# Algorithmic A-Legality: Shorting the Human Future Through AI

Scott Veitch

Faculty of Law

University of Hong Kong

**Abstract**

This article provides a necessary corrective to the belief that current legal and political concepts and institutions are capable of holding to account the power of new AI technologies. Drawing on jurisprudential analysis, it argues that while the current development of AI is dependent on the combination of economic and legal power, the technological forms which result increasingly exceed the capacity of even the most rigorous legal and political regimes. A situation of "a-legality" is emerging whereby the potential of AI to produce harms cannot be restrained by conventional legal or political institutions.

**Introduction**

Progress in artificial intelligence (AI) technology is making law selectively redundant. Already in finance, legal scholars have observed how Big Tech firms are deploying first mover advantage (Pistor 2020) or regulatory arbitrage to "invest first, get approval later" (see Zhang 2024, 32). But this is an old technique, they note, with a pedigree stretching back centuries. This article argues, by contrast, that something new and different is happening which breaks decisively with this logic. The historically unique nature of AI technologies – specifically, the use of autonomous, neural network, machine learning (ML), Large Language Models (LLM) and LLM Agents – signals that an inversion is already under way in which the operational capacities of AI, ML, and LLM Agents cannot be "legalized later". The self-replicating power they generate eludes regulatory accountability, and instead involves a temporal reversal: "legal first, *a-legal* later".

The disruption of law and legal institutions by the rapid development of AI technologies has been signalled in a variety of ways. On the one hand, existing legal concepts such as intention, causation, and liability, struggle to deal with potentially harmful actions of autonomous AI agents. On the other, conventional forms of institutional accountability – political, economic, ethical – are likewise being challenged as to their adequacy. These challenges arise partly as a result of the technologies themselves, specifically their code-based second order capabilities for autonomous learning, decision-making and action. But they also arise as a result of their combination with existing socio-economic forces which are being deployed by certain key actors to secure extensive legal impunity. When this combination reaches a certain stage, it achieves what I call here 'algorithmic a-legality'. This means that autonomous AI starts to move beyond the jurisdictional capacities of contemporary legal orders; beyond, that is, the distinction between legality and illegality.

Algorithmic a-legality has (for now) two main features. First, a kind of 'venture legalism' drives technological growth with the promise of massive returns. Premised on the diminishing value of human resources, legally authorized AI applications create new forms of power and profit that exist at the cutting edge of legal development. Second, through the growing capability, proliferation and embedding of AI in social life and institutions, this new form of algorithmic power starts to selectively exceed the capacity of the original creators, regulators, and the law to hold it to account; writ large, "algorithmic a-legality" amounts to the creation of a global irresponsibility apparatus that overruns existing political and legal modes of holding power to account.

In what follows, I first summarise the idea of "venture legalism" and how this produces the conditions for what I have called "algorithmic a-legality". I then show how the faith in legal regulation of AI is misplaced, partly because of the nature of the new technology and partly because of the more general limitations of modern law and legal institutions. These limitations normally remain unseen or their functions underestimated by advocates of legal regulation and the wider public. The article then explores the implications of these insights for political governance regimes, both democratic and non-democratic. For if, as will be suggested, there is continuity with existing forms of organised impunities, adding AI to the mix takes the irresponsibility apparatus to a whole new level. A novel set of dangers is emerging because neither democratic nor even the most stringent of authoritarian governance regimes is able to constrain the power that is being unleashed in terms that would make it accountable to law or the state. Drawing on insights from jurisprudence – the study of the philosophy and politics of law – the article seeks to explain how and why this is occuring. It concludes by showing how the current trajectory of AI development is putting human communities at risk of large scale, unaccountable, and irreversible harms. This is why it is so important to be clear about the limits of current law and politics in the age of AI. Only by facing their shortcomings can a fresh start be made on rethinking our predicament and revitalising our moribund institutional imaginations.

**From venture legalism to a-legality**

Paul Nemitz, a leading regulation advisor to the European Commission points out three factors that contribute to concerns about the inability or unwillingness to make Big Tech and its AI products accountable. First, AI technologies are being used to sidestep legal regulations that are already in place. "AI systems," he argues "can systematically undermine and circumvent the law. For example, AI systems have modelled tax avoidance, replicated bias in financial lending … and enable anti-competitive product pricing." Secondly, and perhaps more significantly for the future, there is, he writes, a "power asymmetry" between Big Tech and states. Large corporations can easily move across and around borders, while law enforcement "tends to be fragmented across numerous small and under-resourced actors, responsible for discrete functions, operating within separate sectors and siloed by geographical jurisdictions." Enforcement, in other words, even if sought, runs up against real limitations and borders that form the basis of law's jurisdictional capacities. By contrast, since corporations can be agile, then in the absence of international cooperation, states are, ironically, hindered in holding corporations to account by existing structures of legal authority. Finally, many of these same corporations have governments as their clients, supplying them with technologies for core security, surveillance and governance functions. Since many of these functions are targetted by governments *on their own citizens* – biometric and facial recognition systems, for example – governments are themselves reluctant to make publicly available the methods, performance, and outcomes of the technologies they deploy. As a consequence, these corporations are often "shielded from public review by laws protecting trade secrets and establishing investigative privileges." (Nemitz 2023, 251) The cumulative effect of these factors is that corporations and their activities are well protected because much of what they do to escape accountability is already lawfully authorised. In this situation, citizens are doubly disempowered by their governments' and corporations' efforts to keep them in the dark about the full extent and permeation of AI deployment. And where either citizens or investigative activities threaten these arrangements, then, as noted earlier, jurisdictional switching is available to provide a further shield from legal accountability.

As the power and sophistication of AI technology increase, and as they are more thoroughly integrated into the basic fabric of social relations, individual lives, and governance functions in ways that make their withdrawal more and more difficult, then these problems are likely to be amplified rather than solved. It is a measure of the pace of technological development, that any current examples of this are likely to be surpassed reasonably quickly. But to give just one case of how a recent technological development exacerbates the situation, Hu et al (2025) show how decentralized Artificial Intelligence (DeAI) is already disrupting the capacities of regulatory regimes. In technological terms, "the integration of blockchain technologies with advanced AI systems has given rise to on-chain AI agents, which are computational entities that operate autonomously within decentralized infrastructures." (6) The security and privacy provided by such technology can thus be used to screen autonomous operations from oversight or accountability by governments or the public. The radical potential of this lies in what the authors refer to as the performance of "machine sovereignty": "the self-sovereign properties of DeAI, which arise from its technical capacity for operational autonomy and its ability to independently maneuver financial and computational resources." (Hu et al, 6) Given these features, DeAI has the ability to exclude the sovereign power of states and their modes of holding agents to account. Indeed, the ability of states to regulate this new power is inversely proportional to the growth of "machine sovereignty".

As we will see, it does not matter whether or not that state is democratic; what matters is that a new form of power has emerged to compete with traditional state and legislative sovereignty.

We should take note of the sequence here: first, these developments are lawful, or operate at least in the realm of legal loopholes. They combine large financial investment with rapid technological innovation in search of competitive advantage that will foster profitability and economic growth. This is what I have called "venture legalism" (Veitch 2024) and it sees legal actors harness the power of AI for the potential of high value, relatively quick returns. By analogy with "venture capitalism", the term connotes the risky nature of the endeavour and the potential for large gains for the winners. Much of what is going on now by way of securing returns on investment in AI can be described in these terms. But a further stage is already emerging in which the most recent technological developments – such as DeAI – push beyond standard legal coordinates to constitute rival sources of autonomous power.

This is what I call "algorithmic a-legality". It describes a form of power that is no longer amenable to legal categorisation or regulation. Since it influences and guides actors' behaviour, it should be considered as having important normative impact. And yet, as an autonomous, competing, and highly distinctive source of normativity it is immune from being addressed in legal terms. That is, it is not amenable to the application of the most elementary distinction central to the legal order: between legal and illegal behaviour.

It is important to note that the term "a-legality" does not mean *il*legality. As developed in the work of Hans Lindahl, "a-legality" is that which *disrupts* traditional legal categories and so cannot be subsumed within them. He describes it as follows:

"the interruption of legal order wrought by a-legality disrupts the conditions of legal intelligibility of the situation: the act withstands qualification as being simply legal or illegal, and not because it is 'a bit of both', but rather because there is a normative claim that resists both terms of the disjunction, as defined by extant law." (Lindahl 2013, 38)

Illegality, by contrast, means that some behaviour is cognisable by legal categories and it has breached the legal norm; the law applies, and the behaviour contravenes it. By contrast, a case of a-legality is where, for some reason, it is *not possible* to apply the legal norm to the behaviour. Something unusual or peculiar has appeared and it resists being taken into the categories of either

'legal' or 'illegal'. That is why it is not 'a bit of both', but something radically new or different that resists legal comprehension. Christodoulidis explains it in these terms:

"The key argument about 'a-legality' … builds on the introduction of the 'strange', that irrupts in the context of 'normal, unbroken constancy', and is, crucially, an interruption that cannot be domesticated by the order that it puts to question." (Christodoulidis, 2014 951)

It is precisely this inability to be 'domesticated' in the legal system that captures the emerging power of AI. 'Algorithmic a-legality,' as I understand it, describes a new or strange force to which conventional legal categories cannot apply. They cannot get any purchase on this form of power, and so lack the capacity to constrain it. As such, they cannot hold it to standards of legal responsibility.

Let me give an example that might help explain this rather abstract formulation. A case of "venture legalism" would be the behaviour of Big Tech corporations to scrape the internet for data on which to train its increasingly sophisticated generative AI models and thereby advance their digital prowess and profitability. These corporations are legal persons and, by their account, what they are doing is lawful, or least putatively lawful (and no legal finding has yet contradicted this and imposed sanctions for breach of the law). In other words, they are operating lawfully and successfully to increase their market share. Conversely, from the point of view of the holders of copyright whose material – text, music, or images – has been scraped without compensation, the corporations' behaviour is illegal. They have taken something that is not legally theirs and should be prevented from doing so or ordered to provide compensation. From either perspective, the question of the application of the legal standard is clear: the behaviour is either legal or illegal.

A-legality, by contrast, describes the product and impacts of a second-order level of calculation, reasoning, and decision-making operations that emerge from the machine learning capacities of LLMs and LLM agents trained on the data, but which are now autonomous of the human programmers who built the initial models. That power is now genuinely *sui generis*. On the one hand, these operations are no longer produced by legal actors (AI algorithms and LLM agents do not have legal personality) and in that sense their reasoning and decision making processes cannot be subsumed within legal categories (of legal personhood, intention, attribution of responsibility, and so on). Trying to apply legal norms to them amounts to a category mistake. On the other hand, they can shape or have other direct effects on human actors and their behaviour. But the mode of power involved in doing so is likewise not amenable to the application of conventional normativity, such as found in law or morality. The latter set standards of what *ought* to be done, and where a norm is breached by a legal person – by contravening the criminal law, say – then sanctions can follow. Algorithmic autonomy is neither motivated by nor is it susceptable to sanctions. This is why it constitutes (in Christodoulidis's terms) "an interruption that cannot be domesticated by the order that it puts to question." It is thus that algorithmic a-legality – neither legal nor illegal, nor a bit of both, but an immensely potent new form of power –'disrupts' the extant legal order.

If this is correct, what we are witnessing with the rise of autonomous AI is a decline not only in the power of the law, but of the state or sovereign power that issues and enforces the laws. We will come back to this later, but a historical parallel may be helpful in understanding what is going to be at stake. In an influential account of the rise of modernity, Michel Foucault described the emergence and impact of a new form of power – governmentality – that led to the power of states being diminished in terms of their relative importance. This was a form of power that differed qualitatively from the juridical power traditionally exercised by sovereigns. It operated not by "imposing law on men, but of disposing things: that is to say, of employing tactics rather than laws, and even of using laws themselves as tactics – to arrange things in such a way that, through a certain number of means, such and such ends may be achieved." (Foucault 1991, 95) Governments themselves came to rely upon these techniques but unlike traditional sovereign power – which could emanate in legislation, legal adjudication and enforcement – this power was no longer amenable or accountable to conventional legal operations. States were not the source of the power to which they became subject. Instead, the governing functions of states were saturated with something new that could not be controlled by them, not least because they pervaded and came to constitute how governance now worked. The development of new "multiform tactics" – the creation of new knowledges and disciplines and the productive and detailed shaping of conduct through techniques based on these – and not sovereign law was what mattered in the operation of this novel form of power. Conventional understandings of the state – as a unified territory over which a sovereign power ruled through law and legal sanctions – were thus superseded by new techniques of governmentality. The conclusion was stark: the state, Foucault observed, no longer had "this unity, this individuality, this rigorous functionality, nor, to speak frankly, this importance; maybe, after all, the state is no more than a composite reality and a mythicized abstraction, whose importance is a lot more limited than many of us think. Maybe what is really important for our modernity - that is, for our present - is not so much the *etatisation* of society, as the 'governmentalisation' of the state." (ibid, 103, emphasis in original)

An analogous process is happening today with respect to AI. The second-order, autonomous capabilities of AI mean that *its* form of power is neither derived from nor gets its authorisation through law or the state. Its source and mode of operation are independent of law and so it shapes or manages actors and their behviour in ways that are incommensurable with legal normativity. Since its power cannot be subsumed nor held accountable by legal norms or officials it resists conventional legal coordinates both institutionally and conceptually.

But as this new form of power works its way into and throughout society and government institutions – initially because its creators and proponents hold out the promise of efficiency gains, increased competitiveness, and economic growth – it also means that the state's *incapacity* to govern it will become increasingly pronounced. The autonomous power that AI delivers either rewires or, in certain cases, bypasses the state altogether – except in its role as a consumer. Some of the progenitors of this technology have been perfectly candid about this. They have stated clearly their desire to supersede the state in this way, and it is a process that is already achieving success. For example, the CTO of Palantir said in 2021 that its mission was to become "the US government's central operating system", a goal that is being steadily realised during the second Trump administration. As Palantir and similar corporations inveigle themselves into government departments in healthcare, tax, defense, and so on (Medi 2025) then the deployment of AI diminishes and reconfigures the nature of the state itself. The idea that the state is at risk of being overhauled or replaced by new technologies is not at all fanciful. And where many people already believe – or are led to believe – that public institutions are no longer working for the people they are supposedly meant to serve, there is growing openness to the alternatives that AI promises.

As with governmentality as a source of power, the technological capacities associated with 'algorithmic a-legality' have the potential to sideline the importance of the state's sovereign power to make, apply and enforce the law. The widely publicised idea that the danger of AI is that it gets into the *wrong* hands – into the hands of 'bad state' actors – entirely misses this point (perhaps this is the intention of AI's proponents who regularly tout this line). The real danger is that it is transforming all states in ways that deliberately supersede legal, political, and citizens' oversight or control. In deploying a range of applications in surveillance, disinformation, cyberattacks, facial and biometric tracking, identification of threats through data gathering and analysis, and so on, autonomous AI is currently being embedded in how states carry out their basic functions. In the hands of 'good' actors – if we assume democratic states to be such – autonomous AI infrastructures are already being used in military and civilian settings that even now suspend the application of standard legal rules and principles.

In so far as these developments employ forms of AI that exceed legal cognition and control, a new logic of rule comes on board that rivals the state and its institutions. And we are only at the beginning of this. On the one hand, it will reconfigure conventional answers to the question of who or what controls who or what. On the other, while the dangers of bad actors acting badly with AI technology is real enough, so is the risk of 'good actors' acting badly. Yet ironically, because at its own behest, the state has ushered in a technological integration that results in the state itself being downgraded in the hierarchy of power. But it is important to emphasise that authoritarian states are just as vulnerable to this process. Just as the strongest state, when integrated into the global economy, cannot shield itself from domino effects of market crashes; just as the most autocratic jurisdiction cannot stop the wind blowing pollutants over its borders nor unilaterally stop global warming and its effects, so too the with logic of autonomous AI. Existing forms of governance, institutions and structures become data fodder for AI and its agents, to be replicated, manipulated, and controlled by processes over which they no longer have control. It is this mis-match that is the source of new dangers, as the abyss of 'a-legality' starts to widen. Given the technological novelty of digital "machine sovereignty" the modes and consequences of the emerging saturation will have profound effects.

One final factor in all this needs to be highlighted. A core aspect of the context in which these developments are occuring is that of market competition. Whether in East or West, economic growth and competitiveness are central to the aspirations of private and public actors. The desire to increase productivity through AI is explicitly grounded in the promise to reduce the need for or cost of human labour. It is a process, in other words, whose key gambit is betting on the value of human labour declining. It is, to use the markets' term, seeking to "short the future" of human value. The current amalgam of AI and the drivers of the global capitalist economy are undoubtedly unleashing huge amounts of productive power. But this has an entirely predictable downside too, one that is ineradicably built into its operating system, a feature, not a bug. As David Garland explains, while capitalism has both now and in the past unleashed novel forms of creative power and provided modes of decentered distribution in markets, it has always been accompanied with an unrelenting danger:

"there is also a fundamental sense in which capitalism, as a system of economic action, is profoundly *anti-social*. Societies that allow economic life to be governed by the logic of private profit and market competition are societies at risk. They are prone to rapid undirected change; to socially damaging concentrations of wealth and inequality; to crises of accumulation; and to periodic economic collapse – sometimes on a worldwide scale. The chief characteristics of capitalist societies are not stability and equilibrium: these are the unfounded assumptions of economic theory, not facts about real world economies. The chief characteristics of capitalist societies are uncertainty, insecurity, inequality, and undirected change – all of which generate damaging consequences for our social and natural environments." (Garland 2016, 135; emphasis in original)

New AI technologies do not change this; indeed, each of the negative phenomena listed in the quote – instability, crises, and so on – is currently being *exacerbated* by them, including, crucially, serious harm to the natural environment. The reason for this lies in the combination of capitalist economic action with the way in which AI is embedding itself in a constitutive manner in social relations. As Seth Lazar observes: "By constituting the relationships they mediate, algorithmic intermediaries substantially determine our options in those relationships. Everything we can communicate or do to one another via an algorithmic intermediary is made possible by that intermediary. This productive power is pervasive in digital environments, often enabling new kinds of action." (Lazar 2025, 15) So where these new forms of action and relation are hitched to a decentralised, immensely powerful, and (as Garland put it) "profoundly anti-social" economic system, there is every reason to assume that new and far-reaching forms of uncertainty and risk are emerging. Any account of law and politics in the digital age must factor these into its analysis, and it will be defective to the extent that it fails to do so.

**The limits of law**

There is, nonetheless, still a widespread belief in the positive contribution law and legal institutions can make to regulate AI. Despite the challenges posed by new digital technologies, there is a broad consensus about law's capacity to regulate them, since the prevalent question is not whether law *can* but – in the context of a competition to innovate – whether it *should* be used to do so. Striking a balance between innovation and regulation thus assumes that legal regulation *would* work to limit powerful actors and the development and deployment of these new technologies.

The faith being put in law to regulate AI is at one level not surprising. Modern law has achieved an efficacy and social priority through its ability to organise, legitimise, and deploy coercive force in the form of sanctions. It is law that creates powerful actors in the form of officials, agencies, or corporations and holds these and other actors to account through judicial and other enforcement mechanisms. But from a jurisprudential perspective, it is necessary to make some correctives to common perceptions of the role and capacity of conventional law and legal institutions. It is especially important to do this in the context of AI since it brings clearly into view the limitations of current proposals to hold to account new forms of digital power. Put simply, contrary to popular belief law is not nearly so capable a regulatory ally as many would suppose.

Sometimes this is explicitly because law is itself being used by key players to establish a framework for impunity. As we saw earlier, the coincidence of AI and legality is, in such cases, deliberately creating conditions for *non* accountability. Already, in the financial and military sectors, economic, human, and ecological harms are being put beyond the reach of regulatory mechanisms, and this is not by accident. It is the result of more or less concerted efforts by major players in finance, Big Tech, and government to seek to ensure that the harms they cause now and in the future are not attributable to them, even though any benefits – for example in security capabilities or profits – are. Even as powerful a regulatory law-making body as the European Union does this openly, excluding from its jurisdictional reach, for example, the role of AI in "military, defence, and national security" (and regardless of whether the AI is being developed or used by public or private sector actors in these activities). (EU AI Act 2024, Article 2)

But there are deeper challenges to law's regulatory capacities that are usually less apparent. To foreground these, let me draw on a recent important book that addresses the rise and risks of new digital technologies. The problem with them, the author claims, can be succinctly stated: "unaccountable power" (Susskind 2023, 3) The reason this emerges as the key problem is, he argues, because the new medium of power is code, and code operates differently from conventional types of power. Code, writes Susskind, carries "the power to affect how we perceive the world" and thus how we act in it; it is a form of power that is most effective when it operates, as it commonly does, in ways that are invisible to those it affects. The emergence and likely ubiquity of this new type of power mean that "those who write code increasingly write the rules by which the rest of us live. Software engineers", he notes, "are becoming social engineers." (ibid) When these engineers are in turn being propelled by corporate engines whose primary goal – whose primary *obligation* – is enhancing economic value, we are confronted by a rising power that operates at both an individual and global level and contains the potential to organise and manipulate the basic material of social relations.

The question then is what role can law play in this scenario? Susskind argues that it is the purpose of law "to protect us from the predations of the powerful" (ibid, 6), and to do so through providing normative "guiding principles". These include the following: laws can and should be used to preserve the institutions of a free, democratic society; to reduce and regulate unaccountable power while at the same time keeping government intrusion to a minimum; and generally, to "ensure, as far as possible, that powerful technologies reflect the moral and civic values of the people who live under their power." (ibid, 11) These ideas about law's purpose and potential are widely shared, and Susskind's distinctive contribution is to show how they might best be operationalised in the context of what he calls "digital republicanism".

But there are a number of problems with this account of law as a set of guiding principles. It is not the ideal that is problematic; it is instead that there is a blindness in these claims as to what law and legal institutions have done and actually *do* in developed (and developing) capitalist economies. For the legal institutions of a free and democratic society are perfectly compatible with, and have in fact historically *furthered*, precisely the kinds of *unaccountable power* Susskind says they should restrain. And this is not an aberration; it has been the normal operation of modern legal systems in their social and economic contexts. So, to aspire in the digital context to delivering sound results from these principles under existing institutions is not only to dream up too hopeful a future; it is to re-write and then rely upon a fictional past. Susskind shares this "blindness" with most commentators in western democratic societies who seek to use law to hold major forms of power to account. But it is vital to be realistic about what legal institutions actually do, and this means not starting from a belief in a non-existent past but by foregrounding the following three observations about modern law.

First, the assumption that law's primary purposes is to cut down on the use of arbitrary power must be queried. Of course, law and the doctrine of the 'Rule of Law' are generally understood as being opposed to the 'rule of men' where the latter signifies the dangers of unregulated, partial, or unaccountable power. But it was the great proponent of the Rule of Law and the free market, Friedrich Hayek, who pointed out that there is an often unnoticed and unwarranted conflation of law with the Rule of Law's capacity to cut down on arbitrary power: "the belief that so long as all actions of the state are duly authorised by legislation, the Rule of Law will be preserved … is completely to misconceive the meaning of the Rule of Law … The fact that somebody has full legal authority to act in the way he does gives no answer to the question whether the law gives him power to act arbitrarily or whether the law prescribes unequivocally how he has to act." As Hayek concludes, "The law … can legalise what to all intents and purposes remains arbitrary action." (Hayek 1944, 61-62)

This is an important insight, especially in the context of AI. It recognises that law can authorise activities and forms of power that may be entirely *arbitrary* in their operation and effects, and that not only is this perfectly possible, it may have more of a presence than is commonly believed. In fact it has, we might say, a historical provenance, pedigree and continuity that accounts for many harms and much suffering carried out *in the name of the law*. To choose only one of the most serious and enduring examples, consider that up until the last decade of the twentieth century in English law, a married man could rape his wife with no legal accountability whatsoever. This was only the most heinous part of a spectrum of domestic violence by which English law authorised the arbitrary exercise of violence by men over women in a manner that was, for centuries, not amenable to legal review. Now few commentators or officials would doubt that England was a law-bound or Rule of Law society prior to the reform of the law in the 1990s. And yet a large proportion of the population were forced to endure such arbitrary but legalised violence.

In our current context, where AI has and will continue to develop forms of power whose operation and effects are abitrary, there is, as we have seen, every reason to assume that law will not hold such power to account. Indeed, it is in this form – lawful but arbitrary – that 'venture legalism' could be considered to operate: legalising the kind of actions that can cause damage without being liable to legal recourse or further charges of accountability. It is a technique – a legal technology, we might say – that has operated for lengthy periods in so-called 'Rule of Law' societies, and there is little reason to doubt its continuation into the future.

This has a further implication that is crucial for thinking about the ability of law to regulate AI. The difference between law and technological forms of social ordering is not as clear-cut as most observers, particularly legal ones, assume it to be. Many theorists and policy makers argue that unlike law, digital code self-executes in ways that exclude the possibility for rational judgment to intervene. They maintain that the capacity law and legal actors have for reasoned interpretation and deliberation makes it entirely distinct from techological organisation. Drawing on the important work of Mireille Hildebrandt, Diver et al point out the danger of conflating the two in a discussion of the notion of making certain behviour "legal by design". The latter is "a specific type of technoregulation, whereby legal norms are e.g. translated into code or into the design of computing systems such that compliance become automated or semi-automated. Think of self-executing code as in smart contracts or smart regulations, or data-driven techniques for prediction of judgments deployed to make decisions." But in the process of 'translation into code' something crucial happens which the authors contrast firmly with conventional (or text-driven) law, where "legal norms cannot apply themselves and require interpretation, thus enabling contestation". To maintain the contrast between the two means appreciating that "'legal by design' is an oxymoron." (Diver et al 2023, 22)

If this seems intuitively correct, it ignores a major issue. At least in western modernity, the existence, organisation, and application of legal norms always rest upon a condition that certain legal norms are not and *cannot be* contestable. This happens at such a foundational, systemic, level that it is usually never engaged with and is thus frequently overlooked. There are several elements to this. First, constitutional laws and commitments underwrite the substantive and institutional form of a rule of law society in such a way that they cannot be challenged without falling into performative contradiction. Institutionally, a supreme court cannot, for example, question its own jurisdictional capacity to decide whether to answer a legal problem without jeopardising its status *as* a supreme court. Second, the medium in which legal disputes take place and are adjudicated is not open for contestation, and this has profound implications for thinking about freedom and autonomy. The German political philosopher Jurgen Habermas noted the problem precisely: "Subjects who want to regulate their living together by means of positive law are *no longer free to choose the medium in which they can realise their autonomy*. They participate in the production of law only as legal subjects; it is *no longer in their power* to determine which language they will use in this endeavour." (Habermas 1996, 465, emphasis added). So long as the legal system persists, this foundational assumption – and the fundamental laws, institutions, and forms of deliberation that give it expression – are not and cannot be open to contestation because they are *constitutive* of the system itself. In a rule of law society, the ubiquity – the spread and depth – of the legal form thus pervades and underwrites ways of acting and being in the world that *pre*-determine options. As Pierre Bourdieu puts it: "Entry into the juridical field implies the tacit acceptance of the field's fundamental law, an essential tautology which requires that, within the field, conflicts can only be resolved juridically … such entry completely redefines ordinary experience and the whole situation at stake in any litigation." (Bourdieu 1987, 831) People are thus not free to do other than what the law prescribes, and in this sense, it functions exactly like code.

A third element concerns how other normative categories – economic, military, gender-based, or technological for example – latch on to the foundational legal forms to protect certain activities *from* contestation. The examples earlier – legally protected domestic violence, the crises in capitalist markets, or harms in military actions – show how the law can offer impunity for predictable damage. But as long as the rights-based democratic legal institutions of capitalism persist, there can be *no* contestability of the framework rules themselves. Law in this role is thus exactly like 'legal by design': it self-executes in ways that are not open to contestation. And so when it is often said, and rightly so, that AI and other digital technologies operate like a "black box" (Pasquale 2015) – where those subject to its power have no ability to look inside to find out how and why they are being treated or their behaviour directed – it must be emphasised that law too operates as something of a "black box". There are certain foundational assumptions about the operation of the legal system that simply cannot be looked into if the legal system is to continue to exist at all.

What does this mean for the legal regulation of AI? We may grasp the implications by way of posing an objection: Just because the law *can* operate in the ways described, does not mean that it cannot operate *otherwise*. Wasn't the law changed to protect women, for example, from domestic violence? And so, analogously, couldn't law be used to protect citizens from the operation of unaccountable power released by AI? The reply to these questions is straightforward but implacable: such a change may be possible but it would involve jettisoning the constitutive elements of modern law that maintain the capitalist economy: property, rights, individual legal personality, and so on. For it is not only vested interests that limit the role of law. There is a conceptual and institutional apparatus that make it what it is.

As the observations of Habermas and Bourdieu make clear, modern rule of law societies which guarantee the legal protection of persons require that autonomy only be understood as autonomy in law. And this constributes to the second problem with Susskind's account. Modern law does indeed provide the language into which autonomy must be translated if it is to be legally valorised. But when we are dealing with a new form of autonomy of the emerging AI kind, modern law is, as we noted earlier, incapable of gaining any traction on it. It like reading a criminal law statute to a horse. It would make no sense. And the suggestion that granting legal personality or liability to this new form of algorithmic power would be as meaningless as granting it to non-human animals; the latter was abandoned long ago and for good reason. Likewise, granting LLM agents legal personality (as some commentators propose) would only anthropomorphise something that is not capable of bearing it: if a form of power is not amenable to legal regulation then proposing legal solutions will be redundant *ab initio*.

This leads to the third problem with Susskind's analysis. As we saw earlier, he argued that we should seek to ensure that "powerful technologies reflect the moral and civic values of the people who live under their power." Again, this sounds like a good idea. The problem with it, however, is that there *is* no single set of "moral and civic values of the people" in modern societies. It is obvious from even the most perfunctory observation that the reality of modern societies is that they are riven by economic, racial, and gendered divisions in which inequalities are rife and widely acknowledged to be getting worse. To ignore this and to invoke "common" values is little short of an ideological move that seeks to mask the reality. The extremes of wealth and poverty both within particular countries and, more starkly, between countries internationally, make a mockery the claim that there just are "moral and civic values" common to everyone. With levels of deprivation, exploitation, and vulnerability existing so differentially in contemporary societies, the assertion that there are common moral and civic values which can now be applied to AI seems at best ignorant and naïve, and at worst simply cruel. On the other hand, however, it is one of the central roles of modern *law* to cover over and legitimise the facts of inequality and exploitation. This is its ideological function in a capitalist society. By maintaining the principle of the equality of all before the law, the reality of social inequality is both masked and authorised by the assertion that everyone is the holder of equal legal rights. But of course, and as RH Tawney wryly observed, "If the rules of a game give permanent advantage to some of the players, it does not become fair merely because they are scrupulously observed by all who take part in it." (Tawney 1964, p.116)

The reason why this matters in the current context is that where efforts are now being made, even with the best of intentions, to regulate digital technologies by claiming common moral values and equal treatment under existing legal institutions, the outcomes are entirely predictable: legalised inequality and exploitation. Where law regulates power 'normally' in a capitalist society, that is what will be achieved; indeed, that is all that *can* be achieved. Thus ignoring the real conditions and invoking common values of "the people" will merely continue (as Tawney observed) to "give permanent advantage to some of the players". Despite what republican legal and political theory suggests, it is simply not the case that "We are all in this together". To invoke the myth that we are, in the new context of AI will mean business as usual, which is why contemporary law will hold little threat to the key drivers of the AI economy. For these reasons, invoking law as a redeemer of human or humane values in the context of addressing "unaccountable power" is not to propose a common solution at all; it merely offers a smokescreen to both hide and continue the reality of deep social division.

**The limits of democracy**

There is equally an optimism about democratic accountability as a mechanism with respect to the regulation of AI. If the principle of individual equality before the law can co-exist, as we have just seen, with great substantive inequality, does the invocation of collective political equality hold out better hope in the context of AI regulation? The current state of affairs in developed economies suggests some scepticism is also appropriate here. For it is widely acknowledged that states with reliable democratic procedures – regular, free and fair elections, in which all citizens may stand for election in a society with a free press and civil society – are also compatible with great inequalities of power, wealth, and influence. In recent decades, these inequalities have increased (as evidenced by Gini co-efficients) such that the concentration of wealth in the hands of a few continues to expand the gap between rich and poor nationally and internationally.

Big Tech is now playing its role in exacerbating this situation. As well as the visible mega-wealth of a few individuals – Musk, Bezos, and their ilk – there are structural conditions that militate against "people power" holding these individuals and corporations to account. At an elementary level, for example, is no secret that wealthy corporations are able, lawfully, to avoid paying tax on much of their vast profits. The famous democratic slogan that demanded equal citizen participation in the political realm to match their obligations in the fiscal realm – "No taxation without representation" – has been effectively reversed: those wielding the *most* political influence are likely *not* to be paying their fair share of tax: "No taxation *and* super-representation"!

If this is a policy that could be ameliorated, there is again a deeper structural issue that limits the power of collective agency in contemporary polities: the coexistence of democratic procedures and social inequality is characteristically *built into* contemporary democratic states. (John Rawls's *Theory of Justice* is typical in this respect, being explicit about the difference between maintaining equal liberties in the political realm and justifying inequalities beyond that.) Legally and constitutionally, and despite all the rhetoric to the contrary, a democratic polity such as the United States in fact severely curtails democratic rule in one crucial sense. As Karl Polanyi noted, already in the early days of the republic, by separating "the people from power over their economic life" the American Constitution "put private property under the highest conceivable protection and … In spite of universal suffrage, American voters were powerless against owners." (Polanyi 1944, 233) The modern democratic state does not err in its failure to hold economic power to account; that is the democratic state *working*.

When we turn to the ability of democracy or democracies to hold to account the emerging power of AI, it is crucial to understand this profound, constitutive limitation. Given the symbiosis of economic (or market) power and digital technology, then expecting contemporary democratic institutions to be able to keep in check the kinds of power being assembled – on the one hand by Big Tech and on the other, *a fortiori*, by the autonomous power of AI they are developing – is simply delusional. It is *already* the case that democratic oversight of the economy and private power does nothing to reverse the latter's unequalising effects, and this will continue to be so with respect to the power of AI and its corporate chaperones.

To put it another way, there is a "democracy fallacy" that might be added to the "transparency" and "consent" fallacies Edwards and Veale (2017, 23) point out; as they put it, "rights to transparency do not necessarily secure substantive justice or effective remedies. We are in danger of creating a 'meaningless transparency' paradigm to match the already well known 'meaningless consent' trope." Many commentators, academics, and policy makers are falling for this "democracy fallacy", and they will continue to do so despite democracy's undoubted impotence. For the reasons noted, the development of the power of autonomous AI in a capitalist society means it is simply not amenable to meaningful oversight or control in this manner. To be blunt: if democracy is the answer, you've asked the wrong question.

**The limits of authoritarian states**

If democracies can't deal with this type of power, don't authoritarian states have the will and capacity to do so? After all, they have in place well-functioning surveillance and control capabilities over their citizens and market players, and they are able to operate with little or no judicial, media, or civil society supervision. It might indeed be though that AI can only improve the dominance (if not the legitimacy) of these regimes. Yet the heavy reliance of such states on a strong and unaccountable security apparatus marks both a strength and a weakness. And it is the latter that the development of autonomous AI in fact accentuates. For in the longer-term AI will make authoritarian states *more* not less insecure.

There are a number of reasons for this for which there is already evidence. First, the technology itself is mobile. As noted earlier, it can readily evade jurisdictional borders by relocation. No matter how powerful the national security agency is, the global ecology of AI means that the old jurisdictional borders can no longer be taken as watertight. Indeed, authoritarian states *rely* on this very fact to undermine *other* governments and actors beyond their borders. But any state is constantly and increasingly vulnerable to sophisticated hacking, cyber attacks, and misuse by other agents (including increasingly complex and capable LLM agents) which can operate efficiently and effectively at distance and scale.

Second, there is the very real problem of sanctioning. Civil and criminal laws typically deploy the threat of a sanction as a way of guiding citizens' behaviour to be law-abiding. Non-democratic states are particularly well experienced in achieving this in a range of more or less directly coercive ways. Where illegal or otherwise regime-threatening behaviour is suspected, tracking suspects is relatively straightforward and the gathering of evidence and the application of criminal laws – such as sedition, espionage, treason etc – in the interests of national security is all perfectly routine. But where autonomous AI agency develops and is deployed by governments, insecurities rooted in the technology itself are an omnipresent risk not least since efforts at motivating autonomous LLM agents (or "machine sovereignty") are futile. Algorithmic agents do not and will not respond to the kind of reasons humans do. And unlike humans, such agents can relocate, mutate, or proliferate at great speed. They lack entirely the cumbersome materiality of human agents: criminals can be detained and imprisoned but LLMS and their agents cannot be meaningfully put on trial nor in jail. Not even the severest techniques of authoritarian states can threaten or compel these non-human actors into confession or compliance. Since AI agents only offer simulacra of experience, emotion, ethics and so on, they lack entirely the psychological apparatus that motivates human agents to act in certain ways or suffer the consequences for not doing so. In short, autonomous AI agents can undermine the regime, act seditiously, and break the law. But they cannot be punished.

This brings us to the third issue. For there is a crucial reverse side to the observations about the limitations of sanctions: algorithmic agents also lack the ability to be motivated by *positive* attitudes in support of the regime. They lack, for example, the ability to be driven by the virtues that are so central to the ideological successes of the strong state: loyalty, sacrifice, honour, commitment, and so on. If, for instance, loyalty to the nation, or to the Party, or the cause, is a foundational political virtue, it cannot be relied on to motivate algorithmic agents except by way of simulation. But 'machine loyalty' will, in human terms, be no loyalty at all. Despite early efforts to 'align' AI with their values, authoritarian regimes will soon realise their limitations where autonomous technologies develop in unpredictable, unalignable, and unsanctionable ways.

Finally, and added to these technological challenges, the current competition for primacy in the race to lead scientific progress and economic growth inevitably draws authoritarian states into the boom-bust cycles of the global economy. This produces risks that no regime, no matter how powerful, has so far been able to avoid. So where non-democratic states rely on output (or performance) legitimacy in the form of economic growth, rather than input legitimacy (in the form of democratic elections) then failure to secure such growth becomes hazardous for the regime. There is nothing new in this. But the key promise of AI lies in increasing outputs through efficiency and productivity gains that simultaneously reduce the role, cost, and worth of human labour. As such, a contradiction emerges: in the context of diminishing returns on human resources vis-à-vis algorithmic resources, the consequent need to manage social and labour relations will take place under technological conditions that make traditional forms of political management increasingly outmoded as a means of administration. If and when AI delivers on its promised gains, a two-fold lack of control materialises for authoritarian regimes: the fugitive autonomy of AI, and an increase in the number of redundant and disaffected citizens. This is not to say that enemies of the state cannot still be conjured up and harshly treated. But they will be, so to speak, human scapegoats. The real causes of the problem will be far more elusive and non-coerceable.

**Humans in the coop**

Humans have an astonishing ability to make things that they then believe direct their lives. They create gods and then tell themselves it was the gods who created them. They massacre other groups believing that it is what their god or their creed demands of them. What matters in all this is not whether that which is created is real or artificial. What matters is that people act on the belief that these non-human forces actually exist. Historically, the distance between artificial and real has never been clear or particularly meaningful. Likewise with 'artificial' intelligence: it is already having real effects on human societies. To some extent it wouldn't matter whether, as a form of 'intelligence' it really exists or not; as with their gods, what matters is that humans act as if it did.

But to the extent that AI *is* genuinely acquiring autonomous capacities – to reason and to act independently of human impetus or comprehension – then the possibility exists that this time what's artificial *is* for real. Autonomous systems that can create and calculate, judge and perform, in ways that no human is capable of doing, and then act on the basis of these processes without human intervention; all this holds out the possibility of something genuinely new.

I have argued that to the extent that AI acquires autonomy, humans are potentially going to be trapped in systems of action and reasoning that they can no longer regulate: the old tools – law, the state, politics – simply do not have the capacity. Even in their most powerful manifestations, they cannot be relied upon to adequately address or provide remedies for this new human predicament. They lack the technological, coercive, and imaginative capacity to do so.

Ironically, these old tools are currently making things worse, in two senses: they are helping legalise the development of new forms of power that are increasingly pushing beyond human control; and they are complicit in reducing the worth of humanity. Those few who benefit now and in the future from AI are indeed betting on this latter eventuality: that they will make profit from shorting the human future. On this trajectory, the question is not what will save humans, but what will humans be saved *for*? In the main, it will be for two things: their role as consumers and their role as human resources. But there will come to be a third role and this, in a final irony, *will* involve a genuine majority of people: to pick up the pieces when the collapse comes, as it surely will, as it always does. As with the 2008 Global Financial Crisis, the relatively few actors who cause massive problems will be immunised from responsibility, while those who have no causal role will be left to deal with the detritus. The 2008 Crash occurred and the price was paid by the state (ie the taxpayer) because, it was claimed, the banks were too big to fail (in reality amounting to 'capitalism for the poor, socialism for the rich'). Something similar will happen again here: when data resources, platforms, and computing power become so dominant and deeply engrained in social relations that the danger of their collapse will become unthinkable, bail-outs will require that the well-being of the majority be sacrificed because the *technology* is too big to fail.

Yet there is something new this time: given the imminently ubiquitous social dependence on AI technologies, given the intricate and intimate ways in which human and non-human are becoming indissociable, not only will the damage be far greater and more difficult to redress, the capacity and resilience of the people to respond will have been greatly reduced. For that is the logic on which the current trajectory is based. And it is one that laws and states are deeply involved in contributing to, but which in the medium to long term they will be powerless to amend.

This is not far fetched. After all, we are – right now – contemporaries of genocide and planetary destruction, and our sophisticated laws and legal systems do not stop this. If anything, they allow these to happen. Humans, or at least enough of them, continue to put their own and other life systems at risk for the short-term benefit of a few, and all of this is authorised by legal, political, and economic forms of organisation. This is the situation into which autonomous AI is emerging. Given their current complicity and future impotence, it is not at all likely that existing legal institutional forms will be able to temper the situation any time soon. A transformation of imaginative thought is necessary that would match the transformative nature of the technological revolution we are living through. It is not enough – indeed it is a dangerous misconception – to believe that contemporary law and politics will be up to the task when confronted by the growing power and prevalence of AI. For the reasons explained in this article, they won't be. It is time we *began* our thinking with this realisation, rather than insistently trying to resuscitate redundant paradigms.

**Acknowledgements**

The author acknowledges the financial support of the Hong Kong University Grants Council's Humanities and Social Sciences Prestigious Fellowship Scheme (37000422)

Scott Veitch: ORCID: https://orcid.org/0000-0002-8952-4489